\documentclass[4pt]{article}
\usepackage{spconf,amsmath,graphicx,pgfplots,tikz}
\usepackage{caption}

\usepackage{ulem}
\title{Frame Pairwise Distance Loss for weakly-supervised Sound Event Detection}

\name{ \vspace*{-1ex}\shortstack{Rui Tao$^{1, *}$, \thanks{ * These authors contributed equally to this work} 
	Yuxing Huang$^{2,*}$, \\\normalfont
	Xiangdong Wang$^{3}$,
	Long Yan$^{1}$,
	Lufeng Zhai$^{2}$,
	Kazushige Ouchi$^{1}$,   
	Taihao Li$^{2}$ }}

\address{ $^1$ Toshiba China R\&D  Center, Beijing, China.\\
          \{taorui, yanlong\} @toshiba.com.cn, kazushige.ouchi@toshiba.co.jp \\
          $^2$ Zhejiang Lab, Zhejiang, China.\\
          \{huangyx, lfzhai, lith\} @zhejianglab.com \\
          $^3$ Beijing Key Laboratory of Mobile Computing and Pervasive Device, \\
           Institute of Computing Technology, Chinese Academy of Sciences, Beijing, China.\\
          xdwang@ict.ac.cn }

\pgfplotsset{compat=1.18}     
\begin{document}
\maketitle 
\begin{abstract}
Weakly-supervised learning has emerged as a promising approach to leverage limited labeled data in various domains by bridging the gap between fully supervised methods and unsupervised techniques.
Acquisition of strong annotations for detecting sound events is prohibitively expensive, making weakly supervised learning a more cost-effective and broadly applicable alternative. In order to enhance the recognition rate of the learning of detection of weakly-supervised sound events, we introduce a  Frame Pairwise Distance (FPD) loss branch, complemented with a minimal amount of synthesized data. The corresponding sampling and label processing strategies are also proposed. Two distinct distance metrics are employed to evaluate the proposed approach. Finally, the method is validated on the DCASE 2023 task4 dataset. The obtained experimental results corroborated the efficacy of this approach.
\end{abstract}
\begin{keywords}
Sound event detection, multi-branch, metric learning, weakly-supervised learning
\end{keywords}
\section{Introduction}
\label{sec:intro}

Sound carries a large amount of information about our everyday environment and physical events. Therefore, the development of appropriate signal processing methods to extract such information has huge potential in many applications, such as healthcare monitoring, industrial safety, traffic monitoring and so on.
Promoted by the annual DCASE challenges \cite{Turpault2019_DCASE, Turpault2020,Ronchini2021,hershey2021benefit},  SOTA in sound event detection (SED) has progressed rapidly in recent years. Most recent SOTA approaches, e.g., those presented in \cite{Zheng2021,Miyazaki2020,Lin2019,Lu2018}, rely on post-processing \cite{19_Cances}, where neural network performs audio tagging by learning to attend to the time range where the sound event is active. 
The detection capabilities of these systems are actually not ideal and are difficult to reach a practical application level. Therefore, several simpler and more cost-effective  weakly-supervised methods have been proposed
\cite{weak-multi-branch,huang2020multi,weak-guide,weak-mean-teacher, ICME2021-weak,weak-frame-loss,xin22_interspeech, xin23_interspeech}.
such as multi-branch learning \cite{weak-multi-branch,huang2020multi},
guided learning \cite{weak-guide},
mean teacher based method \cite{weak-mean-teacher},
multi-scale gated attention \cite{ICME2021-weak},
multi-instance learning \cite{weak-frame-loss}
audio pyramid transformer \cite{xin22_interspeech},
background-aware modeling \cite{xin23_interspeech}.

In weakly-supervised learning scenarios, integrating metric learning can effectively harness the intrinsic structure of the data, thereby bridging the gap between limited supervision and enhanced model generalization capabilities.
 Many methods have been proposed for metric learning, e.g., Siamese Networks \cite{Siamese-network}, contrastive loss \cite{Contrast-Loss,xu23g_interspeech,su23_interspeech}  and triplet loss 
 \cite{ge2018deep}.
 In recent years, this approach has garnered attention in the field of sound event detection \cite{Contrastive2,Metric-Learning}. 
 To address the challenge of missing labeled samples, the use of pseudo-labeled data has also received considerable attention in the scholarly community \cite{Ebbers2020,Koh2021,22_taorui}.


In this paper, we explore the use of metric learning, named as the Frame Pairwise Distance (FPD) loss,  for training multi-branch \cite{huang2020multi} in the SED model with both Real-world Weakly-labeled (RW) data  and Synthetic Strongly-labeled (SS) data. In order to enable frame-level sampling when the real-world data does not have weakly labels, we use Pseudo-Strong labels generated from Weakly-labeled data (WPS)\cite{22_taorui}. 


The primary objective of our work is to improve the accuracy of SED without involving any additional cost. Our baseline relies solely on weakly labeled data. The obtained experimental results clearly show that the implementation of the proposed methodology could avoid any additional data annotation cost and significantly enhance the recognition accuracy of the system. This improvement could be attributed to the synergistic effects of pseudo labels, strong labels of synthetic data, and the use of the FPD loss function. 

\begin{figure*}
\centering
\includegraphics[width=0.8\textwidth]{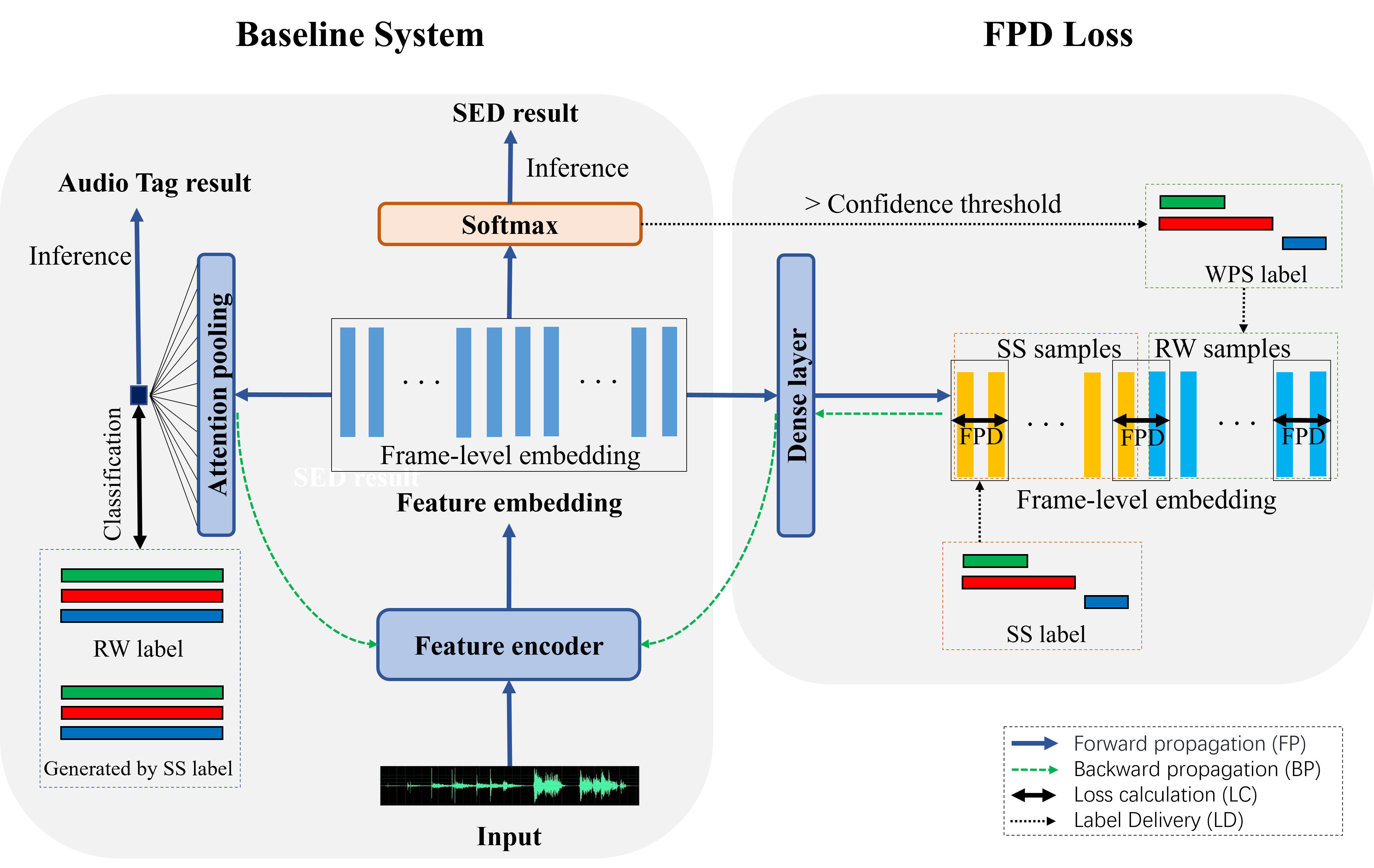}
\caption{ \textbf{An overview of the architecture of the proposed method.} 
This learning scheme requires the computation of two distinct losses during the forward propagation: one is the classification loss based on weak labels, and the other is the FPD loss  based on strong labels. Both losses have their own frame-level embedding, and they collaboratively update a specific feature encoder through back propagation.}
\end{figure*}
\vspace{-0.5em}

\section{METHODOLOGY}
\label{sec:format}

\subsection{Model Overview }
\label{ssec:subhead}

The efficacy of an SED system predominantly relies on its capability to predict sound events at the frame level. Our method emphasizes the modifications in loss calculation at the frame level. As illustrated in Fig.\ 1, the system primarily consists of three branches: a main branch (center), a classification branch (left), and an FPD branch (right).

Within the main branch, both SS and RW data are processed by the feature encoder module to yield frame-level feature embeddings.
The classification branch incorporates an attention pooling module at the embedding level by executing the weakly supervised learning with both SS and RW data. 
For the FPD branch, after the dense layer, the embeddings from these two types of data sources are then paired, and the FPD loss is applied. Its main goal is to refine the feature embedding by ensuring that the distance between two paired frame-level feature embeddings is minimized when they pertain to the same sound event category. 
On the other hand, if the frames belong to different categories,
the distance between their embeddings should be maximized. In order to achieve a consistent distribution of feature embeddings, FPD is also used on pairs derived exclusively from the SS or RW data.
This approach, which is proved to be cost-effective and time-efficient, addresses a critical need in the industrial sector.

\subsection{Label Processing}

For RW data, while weak labels are available, strong labels are absent. To compute the FPD loss, we leverage frame-level embeddings from the classification branch to generate corresponding pseudo labels named as WPS. Thus, the labels employed by the FPD branch comprise both SS labels and WPS labels. In ensuring the quality of WPS, we approached it from three perspectives: firstly, the weak labels employed for the classification branch are of sufficient accuracy; secondly, we established a confidence threshold for utilizing WPS; and thirdly, the model structure of the classification branch is adequately mature.

Conversely, SS data possesses strong labels but lacks weak ones, allowing us to effortlessly derive the corresponding weak labels. Similarly, the classification branch utilizes RW labels and weak labels derived from SS.

\subsection{Sampling Strategy}
\label{ssec:hierarchical_sampling}

\begin{figure}
\includegraphics[width=0.5\textwidth]{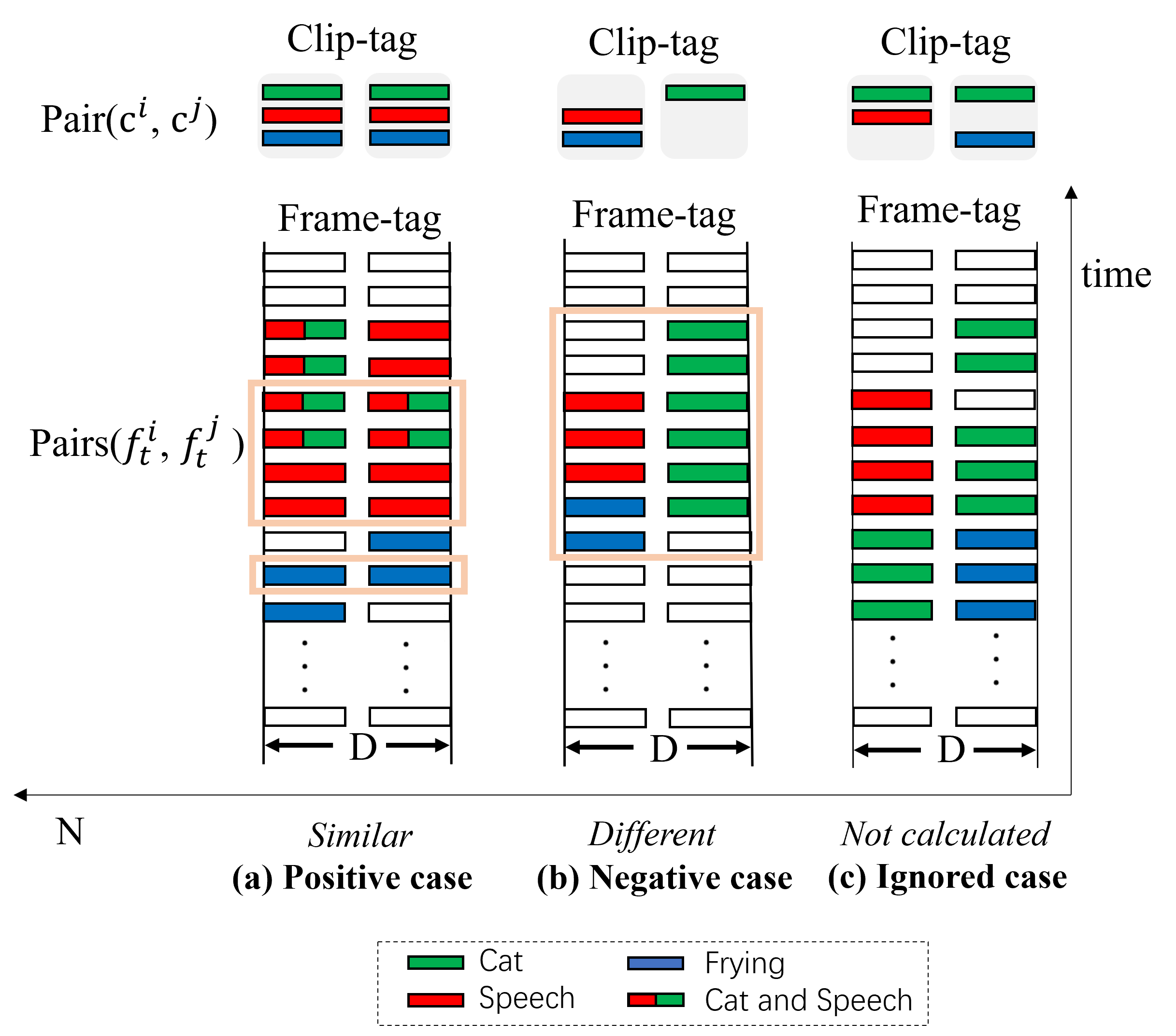}
\caption{ \textbf{FPD loss of Three Sample Cases.} 
The internal logic of the FPD loss is detailed, encompassing the meaning of sample pairs and frame pairs, as well as three types of sampling cases: positive cases, negative cases and ignored cases.}
\end{figure}
\vspace{-0.5em}

The aim of our approach is to place synthetic and real data in a common feature space by treating them equally and leveraging the principles of metric learning to bring frames having similar categories closer and dissimilar ones farther apart. This process involves the reconstruction of the feature space.
 
Our sampling method consists of two phases: curating clip-level pairs from audio clips and then deriving frame-level pairs from these. In a data batch with multiple audio clips, unique combinations are formed, represented as \(Pair(c^i, c^j)\) where \(c^i\) and \(c^j\) are distinct clips.

From these clip-level pairs, frame-level pairs are made by pairing frames from both clips with the same timestamp, represented as \(Pair(f^i_t, f^j_t)\), where \(f^i_t\) is the frame feature of clip \(c^i\) at time \(t\). These frame-level pairs fall into three categories:

\begin{enumerate}
    \item \textbf{Positive Cases:} For clips with matching tags \(c^i\) and \(c^j\), similar frame tags \(f^i_t\) and \(f^j_t\) lead to similar feature embeddings.
    \vspace{-0.5em}
    \item \textbf{Negative Cases:} For clips with different tags \(c^i\) and \(c^j\), dissimilar frame tags increase the distance between their embeddings.
    \vspace{-0.5em}
    \item \textbf{Ignored Cases:} Pairs with neither matching nor completely different tags are disregarded.
\end{enumerate}

\subsection{Distance Metric and Loss Function}

Our approach aims to maximize the distinctiveness of data by increasing the distance between embeddings. We chose such an $\alpha$ value that maximized the distance expansion. The threshold of the increased distance was determined through a training based on the actual data distributions, so as to eliminate the need of manual interventions.

In order to compute the distance metric between frames, we employ both Euclidean (Euc)  similarity  and Inner Product (IP) similarity functions.The loss functions for two given frame vector representations, $A$ and $B$, are defined in the following two subsections.

\subsubsection{Euclidean similarity }

The loss function for Euclidean similarity can be determined as follows:
\begin{align}
D&=  \frac{1}{1 + d}
\end{align}

where, $d$ is the Euclidean distance between vectors $A$ and $B$, which is calculated as follows:

\begin{align}
d&= \sqrt{\sum_{i=1}^{n}(A_i-B_i)^2} \tag{2} 
\end{align}

The loss function for Frame Pairwise Distance (FPD) in terms of the Euclidean (Euc) distance  metric  can be defined as follows:

\begin{equation}
L_{Euc} =\sum_{all\ pairs} \{[-D_{\rm{neg}}+\alpha]_+ + D_{\rm{pos}} ^2\} \tag{3}
\end{equation}
where $[\cdot]_+ = \max(0,\cdot)$. $D_{\rm{neg}}$ is the distance between two negative cases. $D_{\rm{pos}}$ is the distance between two positive  cases, and $\alpha$ is the margin that lies between 0 and 1. Empirical analysis reveals that $\alpha=0.1$ ensures stable model performance.

\subsubsection{Inner Product similarity }

The Inner Product (IP) distance function can be expressed as follows:
\begin{align}
D&= \frac{A \cdot B}{\|A\| \cdot \|B\|} \tag{4}
\end{align}

The loss function for the Frame Pairwise Distance (FPD) in terms of the Inner Product (IP) distance can be expressed as follows:

\begin{equation}
L_{IP} = \sum_{all\ pairs}\{ [-D_{\rm{neg}} + \alpha]_++[D_{\rm{pos}} + \alpha]_+ + L_{norm} \} \tag{5}
\end{equation}

where $[\cdot]_+ = \max(0,\cdot)$, $D_{\rm{neg}}$ is the distance between two negative cases, and $D_{\rm{pos}}$ is the distance between two positive  cases. In order to ensure the normalization of the Inner Product (IP) similarity and align the IP distance, we make them comparable by applying the norm function to frame-level pairs. This results the following:

\begin{equation}
L_{norm} = ||D_{Pair(f^i_t, f^j_t)} + 1 ||^2 \tag{6}
\end{equation}

where, $\alpha$ is the margin that lies between 0 and 1. Empirical analysis reveals that $\alpha=0.1$ ensures stable model performance.

\section{Experiment}

\subsection{Dataset}
Experiments are conducted with the Task 4 benchmark datasets of the DCASE 2023 Challenge, which to DCASE 2019--2023. The task focuses on 10 classes of sound events that represent a subset of Audioset (not all the classes are present in Audioset, while some classes of sound events include several classes from Audioset): Speech, Dog, Cat, Alarm bell ringing, Dishes, Frying, Blender, Running water, Vacuum cleaner, and Electric shaver toothbrush.
\subsubsection{Baseline Data}
 The training datasets contain 14412 unlabeled in-domain training clips (not used in this work), 1578 weakly-labeled training clips with 2244 occurrences  having verified and cross-checked weak annotations, and 2046 synthetic training clips with 6032 occurrences. The baseline model uses only weak labels of the two data.

\subsubsection{Synthetic Data}
The synthetic subset of the development set is generated and labeled with strong annotations using the Scaper soundscape synthesis and augmentation library \cite{salamon2017scaper} tool. For generating additional (potentially infinite) training data, isolated foreground and background sounds are obtained by DCASE Task 4 in combination with the Scaper soundscape synthesis and augmentation library. 
Our method used 2046 synthetic training clips with 6032 occurrences, where the clips were generated in a way to make the distribution per event closer to the  validation set. 

\subsubsection{Evaluation Data}
 The validation set is designed in a way to make the distribution in terms of the clips per class similar to that of the weakly labeled training set. It is the same as the DCASE 2019 task 4 validation set. The validation set contains 1168 clips with 4093 occurrences. It is annotated with strong labels and timestamps (obtained by human annotators). Note that a 10-seconds clip may correspond to more than one sound event.

\subsection{Evaluation Metrics}

Audio tagging F1 score evaluates the accuracy of audio event classifications without temporal precision. 
In contrast, The Event-based F1 metric compares the system’s output and the corresponding reference on an event-by-event basis, measuring the system’s ability to detect the correct event in the correct temporal position. Hence, the Event-based F1 metric acts as an onset/offset detection capability measure-ment scheme. Event-based F1 is the rank index of the DCASE2020 challenge Task 4. 

In summary, Event-based F1 evaluates the precise timing of audio events, segment-based F1 assesses detection within fixed-length segments, and audio-tagging F1 focuses on overall content classification without considering event timings. 
We evaluated our method using these three F1 scores. All experiments were repeated 20 times with random initialization, presenting both average and best results.
While the best outcome can be influenced by randomness, the average F1 score provides a more impartial comparison. 

\subsection{Setup}
In our study, we aim to evaluate our proposed method in comparison with the baseline model. To ensure a fair comparison, we maintain identical parameter settings across both our system and the baseline. The audio is sampled at a rate of 16 kHz. Feature extraction is performed using the short-term Fourier transform coefficients (STFT), utilizing a 2048 sample window and a 255 sample hop size.

Our feature encoder comprises three CNN blocks. Each block is designed with a convolution layer, followed by a batch normalization layer and is subsequently activated using a ReLU layer.

Subsequent to the frame-level embedding, an attention pooling layer is integrated. While this layer accentuates salient features, its core function is to abstract temporal dependencies. This abstraction facilitates the model to concentrate exclusively on event tag extraction from the provided data.

The above Dense layer process is accomplished through a dense projection layer. This layer ingests the feature embedding produced by the feature encoder and outputs a refined embedding representation.

\section{RESULTS AND DISCUSSION}

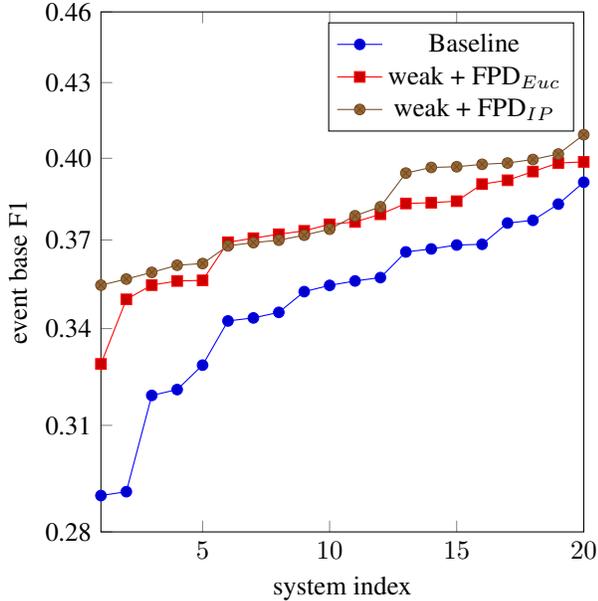
\begin{figure}
\centering
\begin{tikzpicture}
\begin{semilogyaxis}[
height=8.5cm,
width=8cm,
xlabel=system index,
ylabel=event base F1,
xmin=1,
xmax=20,
ymin=0.28,
ymax=0.46,
ytick = {0.28, 0.31, 0.34, 0.37,0.40, 0.43, 0.46},
yticklabels = {0.28, 0.31, 0.34, 0.37,0.40, 0.43, 0.46},
ytick pos=left
]
\addplot coordinates {
(1, 0.289896)
(2, 0.290969)
(3, 0.318984)
(4, 0.320764)
(5, 0.328344)
(6, 0.342512)
(7, 0.343489)
(8, 0.345312)
(9, 0.352222)
(10, 0.35434)
(11, 0.355849)
(12, 0.35698)
(13, 0.3658)
(14, 0.366886)
(15, 0.368267)
(16, 0.368499)
(17, 0.376054)
(18, 0.377031)
(19, 0.382926)
(20, 0.390992)};
\addlegendentry{Baseline}
\addplot coordinates {
(1, 0.328722)
(2, 0.349631)
(3, 0.354413)
(4, 0.355801)
(5, 0.356028)
(6, 0.369215)
(7, 0.370678)
(8, 0.372079)
(9, 0.373256)
(10, 0.375626)
(11, 0.376442)
(12, 0.379209)
(13, 0.383165)
(14, 0.383417)
(15, 0.384028)
(16, 0.390288)
(17, 0.391716)
(18, 0.395032)
(19, 0.398293)
(20, 0.398712)};
\addlegendentry{weak + FPD$_{Euc}$}
\addplot coordinates {
(1, 0.35441)
(2, 0.356477)
(3, 0.358741)
(4, 0.361202)
(5, 0.361785)
(6, 0.367975)
(7, 0.369061)
(8, 0.369935)
(9, 0.371712)
(10, 0.373873)
(11, 0.378682)
(12, 0.381922)
(13, 0.394438)
(14, 0.396555)
(15, 0.396829)
(16, 0.3978)
(17, 0.398243)
(18, 0.399606)
(19, 0.401629)
(20, 0.409123)
};
\addlegendentry{weak + FPD$_{IP}$}
\end{semilogyaxis}
\end{tikzpicture}
\caption{All sorted event-based F1 results.}
\label{fig_whole_event_2019}
\end{figure}

We evaluated three models in our study:
\begin{enumerate}
    \item \textbf{Weak Model:} Our baseline is shown on the left side of Fig.1, which is a weakly supervised learning model.
    \item \textbf{Weak + FPD$_{\rm{IP}}$ Model:} An enhancement of the baseline with the FPD branch leveraging the Inner Product (IP) metric.
    \item \textbf{Weak + FPD$_{\rm{Euc}}$ Model:} A variation of the baseline incorporating the FPD branch with the Euclidean (Euc) distance metric.
\end{enumerate}
The following experiments aim to demonstrate the performance of the Inner Product (IP) distance and Euclidean (Euc) distance on weakly-supervised Sound Event Detection (SED) tasks. We evaluate the effectiveness of these methods using F1 scores at three different temporal granularities, providing a comprehensive assessment of their performance.

\subsection{Event-based Result}
Our findings suggest that integrating FPD leads to an increase of 3.0 points on the average event-based F1 score on the DCASE public evaluation set compared to the baseline. As delineated in Table 1, while the baseline model reached an F1 score of 0.350, the \textbf{Weak + FPD$_{\rm{Euc}}$} and \textbf{Weak + FPD$_{\rm{IP}}$} models recorded scores of 0.374 and 0.380, respectively. 

Fig.\ 3, illustrating the sorted outcomes for 20 iterations of each model. The event-based F1 of \textbf{Weak + FPD$_{\rm{IP}}$ } and \textbf{Weak + FPD$_{\rm{Euc}}$ } are all significantly higher than the baseline model, further substantiates the superior efficacy of FPD-integrated models over the baseline.

\begin{table}[h]
	\begin{center}
 \caption{\label{font-table} Event-based F1 score.}
 \vspace{0.5em}
		\begin{tabular}{|l|c|c|}
			\hline \bf Model & \bf Average F1  & \bf Best F1  \\ \hline
			weak(baseline) & $ 0.350 \pm 0.0281 $ &$ 0.391 $  \\
			weak + FPD$_{\rm{Euc}}$ & $ 0.374 \pm 0.0181 $ &$ 0.399 $  \\
			weak + FPD$_{\rm{IP}}$  & $ \mathbf{0.380 \pm 0.0177}$ &$ \mathbf{0.409} $ \\
			\hline
		\end{tabular}
	\end{center}                                             
\end{table}

\subsection{Segment-based Result}

To garner a comprehensive understanding, we also assessed segment-based and audio tagging F1 scores. For segment-based evaluations, Table 2 reveals that the \textbf{Weak + FPD$_{\rm{Euc}}$} model outperforms with an average F1 score of 0.654, followed closely by \textbf{Weak + FPD$_{\rm{IP}}$} at 0.651, both surpassing the baseline of 0.645.

\begin{table}[h]
	\begin{center}
\caption{\label{font-table} Segment-based F1 score.}
\vspace{0.5em}
		\begin{tabular}{|l|c|c|}
			\hline \bf Model & \bf Average F1  & \bf Best F1  \\ \hline
			weak(baseline) & $ 0.645 \pm 0.0158 $ &$ 0.668 $  \\
			weak + FPD$_{\rm{Euc}}$ & $ \mathbf{0.654 \pm 0.0091 }$ &$ \mathbf{0.673 }$  \\
			weak + FPD$_{\rm{IP}}$  & $ 0.651 \pm 0.0075 $ &$ 0.665 $ \\
			\hline
		\end{tabular}
	\end{center}
\end{table}

\subsection{Audio Tagging Result}

To validate the contribution of the FPD method for weakly-supervised sound event detection tasks, we finally assessed the method using the audio tagging F1 metric. As seen in Table 3, the F1 score for \textbf{Weak + FPD$_{\rm{Euc}}$} reaches 0.713, while for \textbf{Weak + FPD$_{\rm{IP}}$} it is 0.708, both of which significantly surpass the baseline system. The experimental results demonstrate the effectiveness of the FPD method.

\begin{table}[h]
	\begin{center}
\caption{\label{font-table} The Audio tagging F1 score.} 
\vspace{0.5em}
		\begin{tabular}{|l|c|c|}
			\hline \bf Model & \bf Average F1 & \bf Best F1  \\ \hline
			weak(baseline) & $ 0.706 \pm 0.0127$ & $ 0.729 $  \\
			weak + FPD$_{\rm{Euc}}$ & $ \mathbf{0.713 \pm 0.0116 }$ &$ \mathbf{0.741 }$  \\
			weak + FPD$_{\rm{IP}}$  & $ 0.708\pm 0.0089 $  &$ 0.735 $ \\
			\hline
		\end{tabular}
	\end{center}
\end{table}

The peak performances across Tables 1--3 consistently demonstrate the dominance of FPD-enhanced models (\textbf{FPD$_{IP}$} or \textbf{FPD$_{Euc}$}) over the baseline in all evaluation metrics. The effectiveness of the weakly-supervised learning with Inner Product (\textbf{FPD$_{IP}$}) and weakly-supervised learning with Euclidean distance (\textbf{FPD$_{Euc}$}) branches in SED is substantiated by the higher audio tagging F1 scores. These findings underscore the potency and impact of FPD techniques in improving SED systems and emphasize the significance of leveraging  weakly-supervised learning to refine the representation of acoustic signals for better sound event discrimination.

\section{Conclusions}
In this study, we innovatively introduced a Frame Pairwise Distance (FPD) loss branch into a weakly-supervised SED system. By leveraging a minimal amount of synthetic data, the recognition ability of the system was improved. This methodology offers an avenue to augment the applicability of the system without incurring additional annotation costs. Its potential can be extended to other tasks, such as Automatic Speech Recognition (ASR), Computer Vison (CV) and Natural Language Processing (NLP). Our future work would involve validation of the approach on more diverse network architectures and evaluation metrics, notably the Conformer and Wave2Vec 2.0 network structure, and the Polyphonic Sound Detection Score (PSDS) evaluation metric.


\bibliographystyle{IEEEtran}
\newpage
\small
\bibliography{mylib}

\end{document}